\documentclass[twocolumn,aps,showpacs,preprintnumbers]{revtex4-1}
\usepackage{graphicx,latexsym,amssymb,amsmath,color,multirow,mathrsfs,epsfig,bm}
\usepackage{changes}
\usepackage{hyperref}

\hypersetup{colorlinks=true,linkcolor=blue,filecolor=magenta,urlcolor=cyan}

\begin{document}

\title{Origin of octupole deformation softness in atomic nuclei}

 	\author{Bui Minh Loc}
	\email{buiminhloc@ibs.re.kr}
	\affiliation{Center for Exotic Nuclear Studies, Institute for Basic Science (IBS), Daejeon 34126, Korea.}
       \author{Nguyen Le Anh}
	\email{anhnl@hcmue.edu.vn}
	\affiliation{Department of Physics, Ho Chi Minh City University of Education, 280 An Duong Vuong, District 5, Ho Chi Minh City, Vietnam.}
        \affiliation{Department of Theoretical Physics, Faculty of Physics and Engineering Physics, University of Science, Ho Chi Minh City, Vietnam}
	\affiliation{Vietnam National University, Ho Chi Minh City, Vietnam.}
         \author{Panagiota Papakonstantinou}
	\email{ppapakon@ibs.re.kr}
	\affiliation{Rare Isotope Science Project, Institute for Basic Science, Daejeon 34000, Korea}
 	\author{Naftali Auerbach}
	\email{auerbach@tauex.tau.ac.il}
	\affiliation{School of Physics and Astronomy, Tel Aviv University, Tel Aviv 69978, Israel}

 \date{\today}

\begin{abstract}
Recent high-energy heavy ion collision experiments have revealed that some atomic nuclei exhibit unusual softness and significant shape fluctuations. In this work, we use the fully self-consistent mean-field theory to identify all even-even nuclei that are unstable or soft against octupole deformation. All exceptional cases of enhanced octupole transition strengths in stable even-even nuclei throughout the nuclide chart are resolved and the origin is found in basic shell structure. The presence of atomic nuclei exhibiting significant softness to quadrupole-octupole deformation is suggested. These results represent a significant advance in our understanding of the underlying mechanisms of nuclear octupole deformation and have implications for further experimental and theoretical studies.
\end{abstract}


\maketitle

\textbf{Introduction.}
The impact of nuclear deformation on the elliptic and triangle flow measured in relativistic heavy ion collisions was emphasized and formalized in Refs.~\cite{Giacalone2021, Jia2022}.
Subsequently, the analyses published in Refs.~\cite{Zhang2022,nijs2023} of the STAR Collaboration data \cite{starcollaboration2021} provided direct evidence of strong octupole correlation in $^{96}$Zr in heavy ion collisions at very high energy. The question is whether the deformation values reported in low-energy literature are consistent with these new data at high energy. 

The existence of low-lying octupole ($3^-$) vibrations in atomic nuclei, at excitation energies of only a few MeV, was recognized early on Ref.~\cite{LANE196039}. 
They are now among the best-established collective states in nuclear physics and have been observed in almost all even-even stable nuclei throughout the nuclide chart. A review of octupole collectivity can be found in Ref.~\cite{Butler2016}, while Ref.~\cite{Butler1996} offers a review of nuclear reflection asymmetry in general. Experimental information on the first $3^-$ states of even-even nuclides is compiled in Refs.~\cite{Spear1989, Kibedi2002}. Maximal strength values for the transition from the ground state to the first $3^-$ state were observed at neutron numbers $N = 34, 56, 88$, and $134$, and proton numbers $Z = 30, 40, 62$, and $88$ in Ref.~\cite{Spear1990}. Additionally, the numbers $40, 64, 88$, and $134$ were proposed in Ref.~\cite{Cottle1990}. The terms ``octupole-driving particle numbers" \cite{Nazarewicz1984} and ``octupole-magic numbers" \cite{Butler1996} have been used to refer to them.

As already indicated by the theoretical interpretation in Ref.~\cite{LANE196039}, low-lying $3^-$ collectivity in atomic nuclei is driven by the presence of pairs of single-particle states with opposite parity, whose momenta differ by $3$, in the vicinity of and on both sides of the Fermi energy. The presence of such pairs with energy differences much lower than the characteristic shell energy of $1\hbar\omega$ becomes possible thanks to spin-orbit splitting. Enhanced collectivity at lower energies can be expected when the energy difference between such states becomes especially small as a result of the interplay of the spin-orbit coupling strength and the other nuclear interaction terms. Such is the case, for example, of the $2d_{5/2}-1h_{11/2}$ neutron particle-hole pair in $^{96}$Zr. Octupole-magic numbers can be explained in such a way.

Global theoretical analyses of these excitations in even-even nuclei were given in Refs.~\cite{Robledo2011, Agbemava2016, Cao2020, RONG2023137896}. It has been suggested that in order to understand and make reliable predictions about octupole states, one has to go beyond the mean-field approach \cite{Yao2015, Agbemava2016, Barnard2016, Li2016, Nomura2021, Nomura2021_2}. 
However, several questions remain open even then. 
Over time, the exceptional character of some octupole transitions has been revealed experimentally without a satisfactory explanation. The nucleus $^{96}$Zr with $N = 56$ is quite irregular and the various theoretical attempts to study its octupole excitation are inconsistent with one another. As recently discussed in Ref.~\cite{ISKRA2019396}, a variety of theoretical models have been applied to this problem, leading to different results and interpretations as to the origin of the observed transition strength.
Very recently, Ref.~\cite{Spieker2022} reported the enhancement in the light atomic nucleus $^{72}$Se. 

The irregular behavior of the energy of the low-lying $3^-$ in $^{96}$Zr was first found and discussed in Ref.~\cite{Abbas1981}. 
Examining the single-particle spectrum from a mean-field theory perspective, one observed that the $2d_{5/2}$-neutron state is close to being fully occupied, and close by lies the unoccupied opposite-parity $1h_{11/2}$ state. The excitation of each neutron from $2d_{5/2}$ to $1h_{11/2}$, therefore, may form the low-lying $3^-$ state. In contrast, $^{96}$Ru has only $2$ neutrons at the $2d_{5/2}$ and therefore fewer particle-hole states to form the low-lying $3^-$ state, resulting in less collectivity compared to $^{96}$Zr. The presence of this particle-hole pair enhances collectivity and pushes the $3^-$ state to lower energy.

As a collective vibration, the octupole vibration can be studied within the self-consistent mean-field approach, specifically, the self-consistent random-phase approximation (RPA), which is derived as the linearized limit of the time-dependent Hartree-Fock (HF) method~\cite{Co2023}. This was the approach followed early on in Ref.~\cite{Blaizot1976}, where the importance of self-consistency was emphasized. Self-consistent RPA is a unique tool not only for accurately describing collective vibrations and connecting their properties to an underlying energy density functional or effective interaction but also for diagnosing instabilities and broken symmetries: For example, if the self-consistent RPA equations are solved by assuming a spherical ground state, the presence of imaginary solutions in, {\em e.g.}, the quadrupole or octupole channel will indicate that the ground state is, in fact, predicted quadrupole deformed (elongation or compression along the $x$, $y$, or $z$ axis) or octupole deformed (asymmetry along two of the axes), respectively~\cite{Thouless1960, Thouless1961, Rowe1968}.
The spherical ground state is said to ``collapse" \cite{Abbas1981}. Similarly, softness towards shape fluctuations can be expected when there are collective vibrations at very low energy. General arguments with rigorous mathematical treatment on the stability of RPA solutions were presented in Ref.~\cite{Nakada2016}. In addition, as discussed in Ref.~\cite{Nazarewicz1994}, the stability of the RPA is equivalent to the stability condition in the discussion of the Jahn-Teller effect which is an important mechanism of spontaneous symmetry breaking in different fields.

Despite the highly complex phenomena associated with nuclear deformation, we are able to point out the origin of octupole deformation softness in atomic nuclei. The self-consistent mean-field (or RPA) framework is employed here as a diagnostic method to indicate the instability or softness of specific even-even nuclei against variations in the octupole collective variable throughout the nuclide chart. Atomic nuclei which exhibit an extremely high degree of softness to quadrupole and octupole deformation simultaneously are revealed.

\textbf{Method.}
The self-consistent mean-field theory is a powerful tool for studying the properties of atomic nuclei, such as their shapes, energies, and excitation spectra~\cite{Bender2003, RS80, Co2023}. The theoretical foundation of the mean field is provided by the HF theory using the Skyrme interaction, which is one of the most commonly used types of effective interaction. 
RPA is used within the framework of the self-consistent mean-field theory to describe collective motion in atomic nuclei and especially harmonic vibrations. Therefore, this framework is highly suitable for the study of the first $3^-$ octupole state which is low-lying and strongly collective. Self-consistency is ensured when the RPA particle-hole interaction is derived from the same effective interaction used for obtaining the HF ground state. Note that previous research on the $3^-$ octupole state using the Green's function RPA framework \cite{Bertsch1975} was conducted in Refs.~\cite{Abbas1981}, but without full self-consistency.

Computational codes exist for solving the HF and RPA equations and they may include a pairing interaction for describing open-shell nuclei. The publically available computational code described in Ref.~\cite{Colo2013} allows both theorists and experimentalists to perform computations on a wide range of nuclear excitations using fully self-consistent Skyrme HF-RPA, assuming spherical symmetry. 
All relevant terms of the residual particle-hole interaction are incorporated into the calculation, including the Coulomb and spin-orbit terms, the latter being especially relevant for the octupole vibration, as already discussed. 
With pairing correlation included, HF is extended to HF-Bardeen-Cooper-Schrieffer (BCS) and RPA to Quasiparticle RPA (QRPA) accordingly \cite{Colo2021} with the self-consistency maintained. 
We use the above approaches in the present study, i.e., self-consistent RPA and QRPA. The selected Skyrme forces in the present work are SkM* \cite{BARTEL198279}, SLy4, SLy5 \cite{CHABANAT1998231} 
and, for comparison, the SIII \cite{Beiner1975}, which was one of the first Skyrme parameterizations.

The reduced transition strength from the correlated RPA ground state $| \tilde{0} \rangle$ to the first excited state $J^\pi_1$ with the total momentum $J$ and the natural parity $\pi$ is 
\begin{equation}
B(E\lambda) = |\langle J^\pi_1|| \hat{F}_\lambda || \tilde{0} \rangle|^2,
\end{equation}
where $\lambda$ is the multipolarity of the transition, which is $2$ or $3$ for the quadrupole or octupole transition, respectively. For even-even nuclei in the study, $| \tilde{0} \rangle$ is $0^+$, and therefore $J = \lambda$. The electric isoscalar octupole operator is
\begin{equation}
    \hat{F}_{\lambda M}(\bm r) = e \sum_i^A r_i^\lambda Y_{\lambda M} (\hat{r}_i) \frac{1}{2} (1 - \tau_z (i)),
\end{equation}
where $e$ is the electron charge and $\tau_z (i) = -1, 1$ for proton and neutron, respectively. All the transition strengths are presented in Weisskopf units (W.u.).

The degree of stability or softness of the atomic nucleus against density fluctuations, including deformations, is expressed via the polarizability $\alpha_\lambda$, which is obtained from the inverse energy-weighted sum rule of the response function~\cite{RS80, Abbas1981}. In the RPA, it is calculated from the $m_{-1}(\lambda)$ moment
\begin{equation}
    \alpha_\lambda = 2m_{-1}(\lambda)/A,
\end{equation}
with
\begin{equation}
    m_{-1}(\lambda) = \sum_n |\langle J^\pi_n || \hat{F}_\lambda|| \tilde{0} \rangle |^2 E(J^\pi_n)^{-1}.
\end{equation}
The larger $\alpha_\lambda$ (W.u./MeV) is, the less stable the system is against deformations. For example, $\alpha_3$ is around $1$ W.u./MeV in the case of $^{208}$Pb, which is stiff to octupole deformation.

The experimental values are taken from the NuDat database (\href{https://www.nndc.bnl.gov/nudat/}{https://www.nndc.bnl.gov/nudat/}) for the energies and from Table VII of Ref.~\cite{Kibedi2002} for $B(E3)$. A higher value of the strength in W.u. indicates a greater degree of collectivity in the state. In the case of octupole transition strength, $1~\text{W.u.} = 0.0594~A^2 e^2 \text{fm}^6$.

Note that the method can be applied for the diagnostic of quadrupole, octupole, and hexadecapole deformation. Here we focus on diagnosing the octupole softness in atomic nuclei. Quadrupole and triaxial deformation are already known to develop in many mid-shell regions of the nuclide chart \cite{MOLLER20161, ZHANG2022101488}. We will discuss the quadrupole state in the context of quadrupole-octupole softness which is also a type of reflection-asymmetric softness.
\begin{figure}
    \centering
    \includegraphics[width=0.45\textwidth]{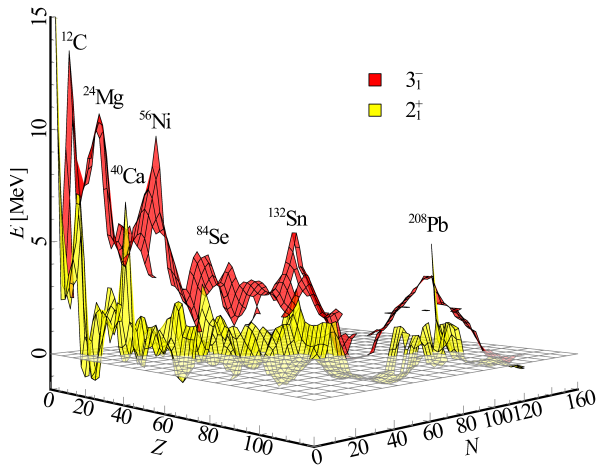}
    \caption{$E(3^-_1)$ and $E(2^+_1)$ obtained from the SLy5 HFBCS-QRPA as a function of proton number and neutron number.}
    \label{fig:3D}
\end{figure}

\begin{figure}
    \centering
    \includegraphics[width=0.45\textwidth]{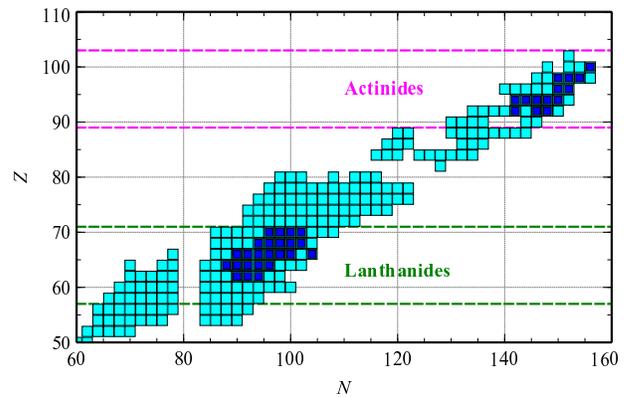}
    \caption{The ``collapses" in the lanthanides and actinides regions based on our diagnostic results using SLy5 QRPA calculation. Cyan: quadrupole collapse; Blue: quadrupole and octupole collapse.}
    \label{fig:CSLanAct}
\end{figure}

\textbf{Results and Discussions.}
We have performed calculations for all stable even-even atomic nuclei with experimental data available and we show our results in Figs.~\ref{fig:3D} and \ref{fig:CSLanAct}. 
A general observation is that there are several regions of quadrupole collapse which is in line with the presence of deformed nuclei in those regions \cite{MOLLER1995, ZHANG2022101488}. Octupole collapse is rarely mentioned in the past. For some atomic nuclei, both quadrupole and octupole cases lead to collapse. These nuclei would be candidates for quadrupole-octupole softness. We stress that this work is \textit{not} about reproducing as well as possible the experimental data for the entire nuclear chart. Instead, we point out the mechanism for octupole deformation softness in atomic nuclei.

\begin{table}[t]
    \centering
    \setlength{\tabcolsep}{6pt}
    \renewcommand{\arraystretch}{1.2}
    \caption{The HFBCS-QRPA results for $^{96}$Zr are irregular in contrast to the results of $^{96}$Ru. The values of $E(3^-_1)$, $B(E3)$, and $\alpha_3$ are in MeV, W.u., and W.u./MeV, respectively.}
    \begin{tabular}{|l|rrr|rrr|}
    \hline \hline 
\multirow{2}{*}{Force}	&	\multicolumn{3}{c|}{$^{96}$Zr} &	\multicolumn{3}{c|}{$^{96}$Ru}	\\ \cline{2-7}
                         & $E$($3^-_1$) & $B(E3)$ & $\alpha_3$ & $E$($3^-_1$) & $B(E3)$ &  $\alpha_3$ \\
\hline
SIII	&	\multicolumn{3}{c|}{collapse} &	$2.198$ & $26.0$ & $1.5$	\\ 
SkM*	&	\multicolumn{3}{c|}{collapse} & 	$3.015$ & $20.7$ & $1.0$	\\
SLy4	&	$0.758$ & $	123.7$ & $	20.3$	& $3.820$ & $24.1$ & $0.8$	\\
SLy5	&	$1.592$ & $56.1$ & $	4.5$	&	$3.685$ & $23.3$ & $0.9$	\\
\hline
Exp.	&	$1.897 $	& 	$53 \pm 6$ &	&	$2.650 $		&	-		&		\\
\hline \hline 
    \end{tabular}
    \label{tab:I}
\end{table}
First, we discuss the difference between $^{96}$Zr and $^{96}$Ru with respect to the octupole excitation.
The recent STAR measurement \cite{starcollaboration2021} showed a significant difference between $^{96}$Zr$+$$^{96}$Zr and $^{96}$Ru$+$$^{96}$Ru collisions that were explained using a transport simulation as large octupole deformation of $^{96}$Zr and large quadrupole deformation of $^{96}$Ru~\cite{Zhang2022}.
Our results for $^{96}$Zr and $^{96}$Ru with different Skyrme forces are presented in Table \ref{tab:I}. We show that a small variation in neutron number can lead to a significant disparity in octupole deformation.

The results of $^{96}$Ru with different Skyrme forces are not dramatically different from the experimental result and from each other and can be considered well-understood. The $^{96}$Ru nucleus has a small octupole polarizability, $\alpha_3 \approx 1$ W.u./MeV as we find, making it hard to become octupole deformed in, for example, isobaric heavy ion collisions. However, note that the result of quadrupole diagnostic for $^{96}$Ru shows the ``collapse" to the quadrupole operator which means $^{96}$Ru is either quadrupole-deformed or soft to quadrupole deformation as is indeed revealed in the relativistic heavy ion collisions \cite{Zhang2022}. 
In reality, the experimental data of $^{96}$Ru show that the $B(E2)$ is small and the ratio $R_{4/2} = B(E2;4^+_1\to 2^+_1)/B(E2;2^+_1\to 0^+_1)$ is close to $2$ rather than $3.3$ which implies $^{96}$Ru is not quadrupole-deformed nuclei in the ground state. The analysis of relativistic heavy-ion collision data in Ref.~\cite{Zhang2022} and our calculation, therefore, suggest that $^{96}$Ru is soft to quadrupole deformation.

In contrast, the results in Table \ref{tab:I} of $^{96}$Zr are irregular. In Ref.~\cite{Abbas1981}, the energy of the $3^-_1$ state of $^{96}$Zr was found to depend very strongly on the single-particle spectrum obtained self-consistently in the framework. When the gap between occupied and unoccupied single-particle energies of opposite parity is abnormally small, the excitation energy of the $3^-_1$ state gets dramatically low. 
For example, the gap between $2d_{5/2}$ and $1h_{11/2}$ given by the calculation with SIII force is so small, that the first $3^-$ state ``collapses". The (Q)RPA calculation is extremely sensitive to this gap. Other earlier (and not self-consistent) (Q)RPA calculations for $^{96}$Zr in Refs.~\cite{Mach1990, OHM1990472, ROSSO199374, FAYANS1994557} with different $2d_{5/2}$-$1h_{11/2}$ gaps, therefore, gave results that are inconsistent with each other.

When the energy of the $3^-_1$ is imaginary in the calculation, we indicate in Table \ref{tab:I} that there is ``collapse". As we assumed a spherically symmetric ground state, the interpretation is that the respective effective interactions predict that the ground state is octupole-deformed. In reality, $^{96}$Zr has the $3^-_1$ state at $1.9$ MeV but our results with the different Skyrme interactions suggest that this nucleus is soft against octupole deformation and a small change in the calculation input can even lead to octupole instability. The SLy5 QRPA calculation reproduces both the excitation energy and transition strength correctly. The octupole collectivity of $^{96}$Zr is $53$ W.u., and the octupole polarizability $\alpha_3$ is $4.5$ W.u./MeV. The large enhancement of octupole collectivity of $^{96}$Zr observed in reactions is understood and reproduced by our calculation within the QRPA framework. Note that in Ref.~\cite{ISKRA2019396}, the reevaluated value of $B(E3)$ for $^{96}$Zr based on consistent results from six independent measurements is $42 \pm 3$ W.u. and the Monte-Carlo Shell-Model calculation gave the value of $46.6$ W.u.. There it was emphasized that the proton contribution is not negligible. We should clarify that, although the instability is driven by the neutron shell structure, both protons and neutrons participate in a collective excitation such as the $3^-_1$ state. Note that QRPA would give us no $B(E3)$ if only neutrons were contributing.

In addition, our results for $^{96}$Mo, which is the isobar nucleus in between $^{96}$Zr and $^{96}$Ru, are $E(3^-_1) = 2.63$ MeV and $B(E3) = 34.8$ W.u with the SLy5 force. The experimental values from Ref.~\cite{Kibedi2002} are $E(3^-_1) = 2.23$ MeV and $B(E3) = 24(3)$ W.u.. Another QRPA calculation with the finite rank separable approximation in Ref.~\cite{Gregor2019} predicted that $E(3^-_1) = 2.95$ MeV and $B(E3) = 34$ W.u..

The condition for softness is not unique to $^{96}$Zr as there are other nuclei with similar characteristics. In the case of $^{96}$Zr, the pair with strong octupole coupling is $2d_{5/2}-1h_{11/2}$. The single-particle spectrum, other single-particle pairs that could drive octupole softness are $2p_{3/2}-1g_{9/2}$, $2f_{7/2}-1i_{13/2}$, and $2g_{9/2}-1j_{15/2}$. They correspond to the octupole-magic nuclei which are around $34, 56, 88$, and $134$ (for the neutron). In addition, other pairs such as $2s_{1/2}-1f_{7/2}$ are also valid. Therefore, we suggest that the smallest octupole-magic number is actually $16$. Table \ref{tab:II} shows the results for selected nuclei with the number of neutrons or/and protons around octupole-magic numbers. Results without pairing (RPA) are also displayed in order to demonstrate that pairing keeps the nucleus less deformed and is essential to reproduce experimental data. However, the presence of pairing does not alter the basic physics we discuss.
\begin{table}
    \centering
    \setlength{\tabcolsep}{3.5pt}
    \renewcommand{\arraystretch}{1.2}
    \caption{The results for selected nuclei with the number of neutrons or protons are around $16, 34, 56, 88$, and $134$ (neutron only). Enhanced octupole transitions are found from light to heavy octupole-magic nuclei, while ``collapse" may be obtained. The values of $E(3^-_1)$, $B(E3)$, and $\alpha_3$ are in MeV, W.u., and W.u./MeV, respectively.}
    \begin{tabular}{|cl|rrr|rrr|}
    \hline\hline
	 Nuclei & Force & 	$E$($3^-_1$)  &	$B(E3)$ & $\alpha_3$ & $E$($3^-_1$) & $B(E3)$ & $\alpha_3$ \\
        &  & \multicolumn{3}{c|}{RPA} & \multicolumn{3}{c|}{QRPA} \\ \hline 
        $^{32}_{16}$S$_{16}$ & SkM*	&	$5.515$ & $16.5$ & $1.0$ &  $5.492$ & $16.6$ & $1.0$ \\
        & SLy4	&	$6.095$ & $19.4$ & $1.0$ & $6.077$ & $19.5$ & $1.1$ \\
        & SLy5	&	$6.216$ & $19.9$ & $1.0$ &   $6.197$ & $20.1$ & $1.1$ \\
        & Exp.	&	 &	&	&  $5.006 $ & $30 \pm 5 $ &	\\ \hline
        $^{64}_{30}$Zn$_{34}$ & SkM*	&	$1.959$	& $5.8$ & $1.4$ & $3.315$ & $18.5$ & $1.4$ \\
        & SLy4	&	$3.381$ & $13.3$ & $1.2$  & $4.243$ &	$24.7$ & $1.2$ \\
        & SLy5	& $3.431$	 & $13.6$ & $1.2$ & $4.265$ & $24.8$ & $1.2$  \\
        & Exp.	&	 & & 	& $2.999 $ & $20 \pm 3$ &	
            \\ \hline
        $^{72}_{34}$Se$_{38}$ & SkM*	&	$0.974$ & $61.7$ & $7.9$ &  $0.958$ & $91.5$ & $11.9$ \\
        & SLy4	&	$2.001$ & $47.0$ & $3.1$  & $2.305$ & $52.3$ & $3.0$ \\
        & SLy5	& $1.862$ & $46.6$ & $3.4$ &  $2.312$ & $50.0$ & $2.9$\\
        & Exp.	&	 & &	&  $2.406 $ &	$32 \pm 11$	& \\ \hline
        $^{98}_{40}$Zr$_{58}$ & SkM*	&	\multicolumn{3}{c|}{collapse}  & \multicolumn{3}{c|}{collapse}  \\
        & SLy4	&	\multicolumn{3}{c|}{collapse} & \multicolumn{3}{c|}{collapse} \\
        & SLy5	&	\multicolumn{3}{c|}{collapse} & $1.199$ & $74.1$ & $8.1$ \\
        & Exp.	&	& &	&  $1.806 $ & -  &\\ \hline
        $^{146}_{\phantom{1}56}$Ba$_{90}$ & SkM*	&	\multicolumn{3}{c|}{collapse} & \multicolumn{3}{c|}{collapse}  \\
        & SLy4	&	\multicolumn{3}{c|}{collapse} & $1.604$ & $42.8$ & $3.2$ \\
        & SLy5	&	\multicolumn{3}{c|}{collapse} & $1.444$ & $48.7$ & $3.8$ \\
        & Exp.	&	&	&	& $0.821 $  & $48^{+21}_{-29}$ & \\ \hline
        $^{152}_{\phantom{1}62}$Sm$_{90}$ & SkM*	&	\multicolumn{3}{c|}{collapse} & \multicolumn{3}{c|}{collapse} \\
        & SLy4	&	\multicolumn{3}{c|}{collapse} & \multicolumn{3}{c|}{collapse} \\
        & SLy5	&	\multicolumn{3}{c|}{collapse} & \multicolumn{3}{c|}{collapse} \\
        & Exp.	&	&	&	& $1.041$  & $14.2$ & \\ 
        \hline
        $^{226}_{\phantom{1}88}$Ra$_{138}$ & SkM*	& \multicolumn{3}{c|}{collapse} & \multicolumn{3}{c|}{collapse} \\
        & SLy4	& \multicolumn{3}{c|}{collapse} &  $0.966$ & $74.7$ & $5.4$ \\
        & SLy5	& \multicolumn{3}{c|}{collapse} & $1.162$ & $61.4$ & $3.7$ \\
        & Exp.	& &	 &	& $0.322 $ & $54 \pm 3$ & \\ \hline 
        $^{240}_{\phantom{1}94}$Pu$_{146}$ & SkM*	&	\multicolumn{3}{c|}{collapse} & \multicolumn{3}{c|}{collapse} \\
        & SLy4	&	\multicolumn{3}{c|}{collapse} & \multicolumn{3}{c|}{collapse} \\
        & SLy5	&	\multicolumn{3}{c|}{collapse} & \multicolumn{3}{c|}{collapse} \\
        & Exp.	&	&	&	& $0.649$  & $17.1$ & \\ \hline
        \hline
    \end{tabular}
    \label{tab:II}
\end{table}

Tables \ref{tab:I} and \ref{tab:II} show that the results for octupole-magic nuclei are extremely sensitive to the choice of Skyrme force. While SLy4 and SLy5 functionals give very similar results in many nuclear structure studies, they are distinguished in the case of octupole deformation softness. The difference between them is in the terms which depend on the spin-orbit densities \cite{CHABANAT1998231}. The spin-orbit interaction plays a key point in the single-particle spectrum which, as we saw, largely determines the octupole-magic numbers. A small change in the spin-orbit component makes a significant change in the calculated $3^-_1$ octupole state.

The result of the calculation with the SLy5 is interpreted as follows. First, experimental low-lying octupole states in atomic nuclei without strong octupole correlation are reasonably reproduced by the spherical QRPA framework, but for soft-octupole deformed nuclei like $^{96, 98}$Zr, different Skyrme forces yield varying results. The high sensitivity to the input and the occasional collapse means that the spherical QRPA calculations can diagnose nuclei soft to octupole deformation. Second, assuming a spherical shape remains reasonable for $^{96, 98}$Zr. With SLy5 force one can construct a stable ground state and reproduce experimental data well.
However, when considering atomic nuclei that display extreme softness to octupole deformation, the spherical QRPA calculation exhibits a ``collapse", as observed in cases like $^{152}$Sm and $^{240}$Pu (Table \ref{tab:II}).
These are candidates for nuclei with the octupole shape in the intrinsic frame. Ubiquitous quadrupole-deformed nuclei throughout the nuclide chart complicate the situation as discussed in Refs.~\cite{Metlay1995, Mueller2006, Robledo2013}. Nevertheless, the spherical QRPA calculation can still diagnose nuclei soft to octupole deformation. Atomic nuclei shown in Table \ref{tab:II} are discussed in the following.
 
The nucleus $^{32}$S is a double-octupole magic nucleus ($N = Z = 16$). The fully occupied state $2s_{1/2}$ is strongly coupled with the unoccupied state $1f_{7/2}$. The value of $E(3^-_1)$ is not imaginary for such light nuclei, but there is the enhancement of octupole collectivity. The experimental value of $B(E3)$ for $^{32}$S is $30$ W.u. according to Ref.~\cite{Kibedi2002} making it the strongest known $B(E3)/A$ value. The most recent value is $16 \pm 3$ W.u. in the evaluation of Ref.~\cite{OUELLET20112199} (page 2265). Our result is $20$ W.u. with the SLy5 force.

Reference~\cite{Spieker2022} reported a recent experiment that showed a notably higher octupole strength of approximately $32$ W.u. for $^{72}$Se. The origin of enhanced octupole strength is well-explained in our discussion. The positive parity $1g_{9/2}$ state comes close to the negative parity states $2p_{3/2}$ and $1f_{5/2}$. It triggers the enhancement of the octupole transition strengths in atomic nuclei with the number of protons or neutrons equal to $32-38$. Note that many nuclei in this region are known to be quadrupole deformed. While a spherical calculation can reveal the enhancement of the octupole transition, it cannot precisely reproduce experimental values for all nuclei in the quadrupole-deformation region. Some reasonable results for $^{64}$Zn, $^{72}$Se, and $^{98}$Zr are shown in Table~\ref{tab:II}. Note that in $^{84}_{34}$Se$_{50}$, the effect of octupole deformation softness is surpassed by the stiffness of the closed shell structure as $N = 50$ (see Fig.~\ref{fig:3D}).

The results for $^{146}$Ba, $^{152}$Sm, $^{226}$Ra, and $^{240}$Pu are presented in Table \ref{tab:II} as examples for nuclei with both $Z$ and $N$ being around octupole-magic numbers, $Z \approx 56$ and $N \approx 88$ or $Z \approx 88$ and $N \approx 134$ (double-octupole magicity). These nuclei have been the subject of numerous studies \cite{Nazarewicz1984, Gaffney2013, Bernard2016, Bucher2016, Bucher2017, Xia2017, Dobaczewski2018, Spieker2018, Butler2019, Butler2020, Chishti2020}. Our results in Fig.~\ref{fig:CSLanAct} provide an overall picture of these regions.

The double-octupole magicity may lead to an octupole shape in the intrinsic frame as has been suggested before in the lanthanide ($Z = 57-71$) and actinide ($Z = 89-103$) region. For example, Ref.~\cite{Ebata2017} found $28$ atomic nuclei that have this property. We remark that in our calculations with the SLy5 force, the ``collapse" related to octupole deformation occurs in the following even-even nuclei $^{152-156}$Sm, $^{152-160}$Gd, $^{156-170}$Dy, $^{162-170}$Er, and $^{166-172}$Yb in the lanthanides, and $^{234,238,240}$U, $^{236-244}$Pu, $^{246,248}$Cm, $^{248-252}$Cf, and $^{256}$Fm in the actinides. The result indicates nuclei with strong octupole correlation or extremely soft against the octupole deformation. The strong octupole correlation in $^{240}$Pu and $^{238}$U were shown in experiments in Refs.~\cite{Wang2009,Zhu2010}. Note that the experimental data suggest no static octupole deformation for the ground states of these two nuclei.

Combined with our diagnostics for the first $2^+_1$ quadrupole excitation (see Figs.~\ref{fig:3D} and \ref{fig:CSLanAct}), the simultaneous ``collapse" for quadrupole and octupole deformation is also found for $38$ nuclei. In Fig.~\ref{fig:CSLanAct}, they are shown as blue squares. These $38$ nuclei are, therefore, candidates for quadrupole-octupole-deformed nuclei which can enhance some unusual phenomena, such as observations of violation of fundamental symmetries (invariance with respect to coordinate inversion and time reversal)~\cite{Auerbach1996, Spevak1997}.

\textbf{Conclusions and outlooks.}
We have utilized a simple method based on the fully self-consistent mean-field (RPA) framework to diagnose octupole softness in atomic nuclei. We have highlighted the role of spin-orbit splitting. An interesting question is whether the evolution of shell structure towards the boundary of nuclear stability (the drip line) results in new octupole magic numbers and reflection asymmetry in exotic nuclei. Experimental information on the strong octupole correlation in atomic nuclei could be used to constrain the spin-orbit splitting, and in general, the energy density functionals. A modern approach that considers the interaction between quadrupole and octupole modes beyond RPA is necessary for quantitative investigations into unusual phenomena, such as violations of fundamental symmetries. Maintaining self-consistency throughout the analysis is crucial.

\begin{acknowledgments}
We would like to thank Prof. Vladimir Zelevinsky for the discussion on quadrupole-octupole deformation and Prof.~Jiangyong Jia for the discussions on relativistic heavy ion collisions.
B.M.L. was supported by the Institute for Basic Science (IBS-R031-D1). N.L.A. acknowledges the support of the VNU-HCM post-graduate scholarship program. P.P. was supported by the Rare Isotope Science Project of the Institute for Basic Science funded by the Ministry of Science, ICT and Future Planning and the National Research Foundation (NRF) of Korea (2013M7A1A1075764).
\end{acknowledgments}

\bibliography{refs}

\end{document}